\documentclass[runningheads]{svmult}

\usepackage{makeidx}   
\usepackage{graphicx}  
\usepackage{subeqnar}  
\usepackage{multicol}  
\usepackage{physmubb}  
\makeindex             

\begin{document}
\title*{Recent Results from STAR}
\toctitle{Recent Results from STAR}
%
%
\titlerunning{Recent Results from STAR}
%
\author{Markus~D.~Oldenburg\thanks{Max-Planck-Institut f\"ur Physik, Munich,
    Germany} for the STAR Collaboration}

\authorrunning{Markus~D.~Oldenburg for the STAR Collaboration}
%
%

\maketitle              

\section{The Relativistic Heavy-Ion Collider}
The Relativistic Heavy-Ion Collider (RHIC) \cite{rhic} is situated at Brookhaven
National Laboratory (BNL) on Long Island, New York. It has operated since 2000
and delivers protons to Gold (Au) nuclei at a center of mass energy of up to
$\sqrt{s_{\mathrm{pp}}} = 500\,\mathrm{GeV}$ for protons and
$\sqrt{s_{\mathrm{NN}}} = 200\,\mathrm{GeV}$ for heavy-ions. The two independent
accelerator rings have circumferences of 3.8\,km.

During the last run period starting in fall 2001 the accelerator was filled with
55--56 bunches of Au ions per ring. Each bunch contained on the average
$7.5\times 10^8$ Au nuclei at a center of mass energy of 200\,GeV per nucleon
pair. The peak luminosity was $5\times
10^{26}\,\mathrm{cm}^{-2}\,\mathrm{s}^{-1}$.  In addition we had a very
successful polarized proton run at a storage energy of 100\,GeV, where RHIC
delivered $0.8\times 10^{11}$ protons in 55 bunches per ring at a peak
luminosity of $1.5\times 10^{30}\,\mathrm{cm}^{-2}\,\mathrm{s}^{-1}$. The
polarization was only around 25\,\%, but we are looking forward to a major
improvement in this years run.  \cite{drees}

At its 6 collision regions RHIC hosts 4 dedicated heavy-ion experiments: Two
smaller ones called BRAHMS and PHOBOS, and the two large experiments PHENIX and
STAR. In addition the pp2pp experiment will study elastic p-p scattering
at high energies. 

\section{The STAR Experiment}
\begin{figure}
\begin{center}
\includegraphics[width=.95\textwidth]{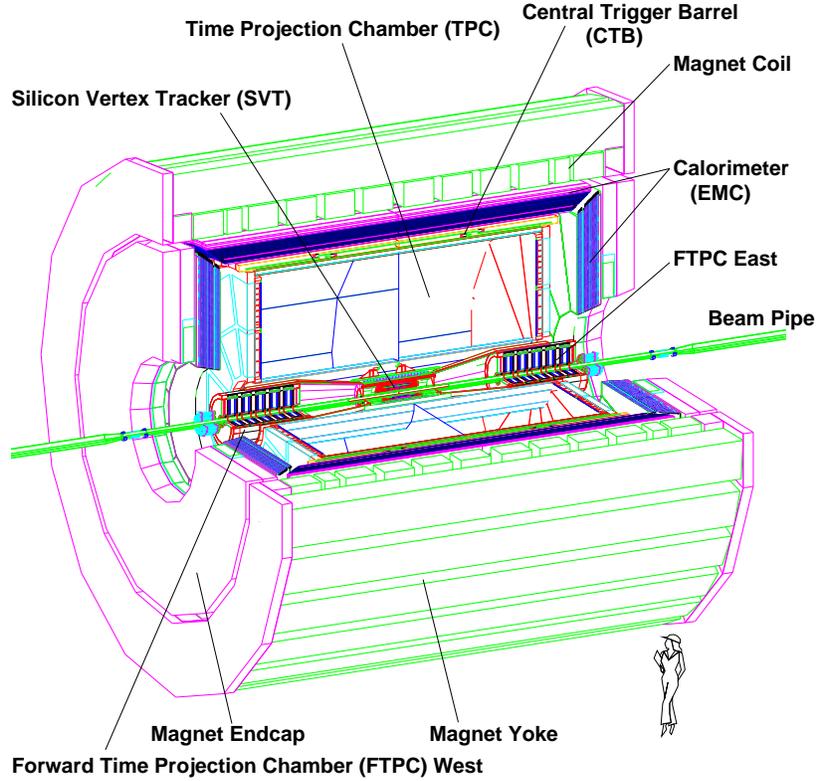}
\end{center}
\caption[The STAR detector]{The STAR detector}
\label{star_pic}
\end{figure}

STAR focuses on measuring hadronic observables over a wide rapidity range and
into a large solid angle. An overall layout of STAR can be seen in
Fig.~\ref{star_pic}. The main detector of STAR is the world's largest time
projection chamber (TPC) \cite{tpc}. A silicon drift detector (SVT) surrounds
the interaction vertex. The high (pseudo)rapidity range is covered by two
Forward-TPCs (FTPCs). These detectors, which use a radial drift field
perpendicular to the applied magnetic field, were developed and built at the
Max-Planck-Institut f\"ur Physik in Munich \cite{ftpc}. Inside the magnet sits
an electromagnetic calorimeter (EMC), which will be fully operational in future
runs.

The central trigger barrel (CTB) covers the region around the TPC. It detects
particle multiplicity. A pair of calorimeters (ZDCs) at zero degrees with
respect to the beam and about 18\,m away from the nominal collision point
measures neutral energy. In combination these two detector types, CTB and ZDC,
allow event centrality (or impact parameter) selections.

\section{Measurements of Anisotropic Flow}
Flow produces the anisotropic shape of the transverse momentum distribution in
non-central heavy-ion collisions. For its measurement a Fourier decomposition of the
momentum distribution with respect to the reaction plane \cite{flow} is
performed. The reaction plane is given by the beam axis and the closest distance
of the centers of the two colliding nuclei. For each event we calculate the
Fourier coefficients $v_n$ for different harmonics $n$:
\begin{equation}
v_n = \langle\cos n\!\cdot\!\left(\phi-\Psi_n\right)\rangle\; ,
\end{equation}
where $\phi$ denotes the transverse emission angle of one particle,
\begin{equation}
\phi = \tan^{-1}\frac{p_y}{p_x}\; ,
\end{equation}
and the reaction plane angle $\Psi_n$ is estimated on an event-by-event basis
by searching for the maxima of the $\phi$-distribution.

The $2^\mathrm{nd}$ Fourier coefficient $v_2$ is the so called elliptic flow.
The ellipticity in spatial coordinates comes about because of the asymmetric
overlap of the two colliding nuclei in a non-central collision.  However, a
finite cross-section for re-interactions is required for this initial state
spatial asymmetry to translate itself into a final state momentum space
asymmetry; and the cross-sections must be large at early times otherwise the
details of the initial collision geometry could not manifest themselves in the
final state momentum distributions.

We see elliptic flow in pion, kaon, proton and lambda spectra. $v_2$ is large
and in good agreement with hydrodynamic models for ultra-relativistic heavy-ion
collisions \cite{first_flow_paper}. Even the mass dependence is well described
by hydrodynamics \cite{mass_dependence_flow1,mass_dependence_flow2}. This is
interesting because these models \cite{hydro} assume local thermal equilibrium
and large amounts of elliptic flow can only be generated if equilibrium is
established very early in the time history of the collisions.

\begin{figure}
\begin{center}
\includegraphics[width=.95\textwidth]{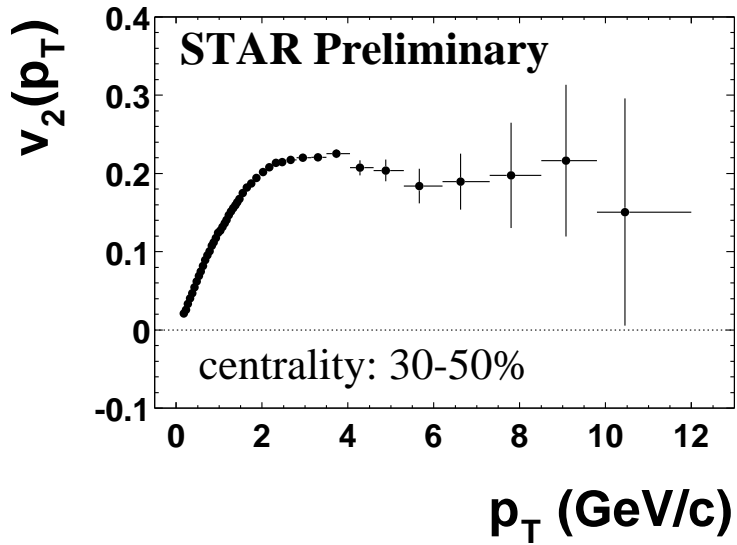}
\end{center}
\caption[Elliptic flow $v_2(p_\mathrm{T})$]{Elliptic flow $v_2$ as a function of
  $p_\mathrm{T}$. At high $p_\mathrm{T}$ the signal saturates which is in
  qualitative agreement with model calculations \cite{jet_quenching_model}
  assuming partonic energy loss. However, this model implies unnaturally high
  gluon densities at mid-rapidity}
\label{flow_pic}
\end{figure}

One surprising thing about our data is that elliptic flow persists to high
momentum and then it saturates, see Fig.~\ref{flow_pic}. This applies for
collision energies of $\sqrt{s_\mathrm{NN}}= 130\,\mathrm{GeV}$ as well as for
$200\,\mathrm{GeV}$ and for all centralities. The flow pattern follows the
prediction of hydrodynamics up to about 2\,GeV/c but then hydro predicts that
the curve should continue to rise.  In contrast we observe that the elliptic
flow saturates. Thus, there is some mechanism even at high $p_{\mathrm T}$ which
allows the asymmetry in the emission pattern to persist to the highest measured
transverse momenta.  This mechanism may be partonic energy loss or another
exotic process. However, at $p_\mathrm{T}>6\,\mathrm{GeV/c}$ non-flow effects
could have a considerable contribution to the observed $v_2(p_{\mathrm{T}})$.

In a model \cite{jet_quenching_model} which adds to a perturbative QCD
calculation a parameterized hydro component, a similar behavior was observed
\cite{high_pt_flow}. There the value of $v_2(p_{\mathrm{T}})$ at saturation was
found to be sensitive to the initial gluon density. This is a hint that the
saturation is due to an energy loss of the partons in the dense medium.

\section{Jets in Nucleus-Nucleus Collisions}
Jets are hard to find in heavy-ion collisions because the high multiplicity of
particles hides the jets and the calorimeter (EMC) was not completely installed
for data taking in 2001. Therefore the question how jets in Au-Au collisions
compare to jets in p-p collisions was attacked using a statistical correlation
technique.  The correlation function is built by identifying a particle with a
transverse momentum that exceeds a trigger threshold and then looking for
associated high-$p_\mathrm{T}$ particles at similar angles and rapidity
intervals:
\begin{equation}
C_2\left(\Delta\phi,\Delta\eta\right) = \frac{1}{N_\mathrm{trigger}}\frac{N\left(\Delta\phi,
  \Delta\eta\right)}{\mathrm{Efficiency}}\; .
\end{equation}

The correlation data show that jets exist in central collisions at RHIC and we
have previously reported on them at 130\,GeV \cite{high_pt}. However, other
correlations exist in heavy ion collisions which may not be present in p-p
collisions. Elliptic flow is an example. We need a technique to distinguish
one correlation from the other.

\begin{figure}
\begin{center}
\includegraphics[width=.95\textwidth]{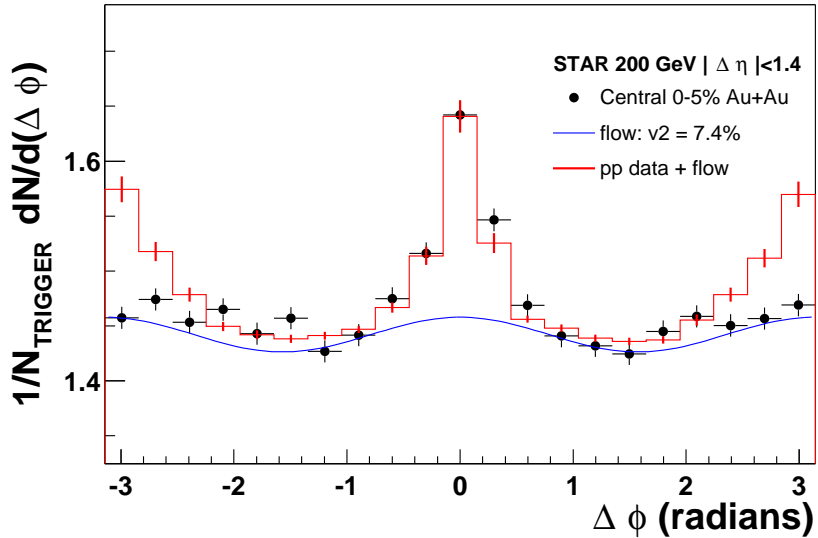}
\end{center}
\caption[Suppressed jets in backward going directions]{The observed jet like
  correlation function in heavy-ion collisions is compared to the correlations
  seen in p-p collisions. The effects of elliptic flow are added to the p-p
  reference function. Note that the correlation is suppressed at $180^\circ$ in
  Au-Au collisions}
\label{jets}
\end{figure}

We have done this by looking in the non-jet region $(45^\circ <
\Delta\phi<125^\circ)$ and building up the correlation function in and out of
the jet cone. The new correlation function is shown below:
\begin{equation}
C_2\left(\mathrm{Au}+\mathrm{Au}\right) = C_2\left(\mathrm{p}+\mathrm{p}\right) +
A\cdot\left(1+2v^2_2\cos\left(2\Delta\phi\right)\right)\; . 
\end{equation}
It includes the p-p correlation and the effects of elliptic flow. The $v_2$
parameter was determined independently by a reaction plane analysis and the
magnitude, $A$, of the flow term was fit in the non-jet cone region, see
Fig.~\ref{jets}.

The correlation functions for Au-Au and p-p (+ flow) collisions are very similar
in the forward jet cone regions $(|\Delta\phi|< 1\,\mathrm{radian})$. But the
correlation functions for central collisions are different in the backward cone
region and, in particular, the backward going jets in Au-Au are suppressed.

It is possible that a jet can only be seen when it originates near the surface
of the collision zone. The backward going jet could be suppressed either because
the fireball is opaque to high-$p_\mathrm{T}$ particles, or, perhaps, the
angular correlation with the away side jets is destroyed by multiple gluon
exchange in the gluon saturated core of the fireball.

\section{Ultra-Peripheral Heavy-Ion Collisions}
In ultra-peripheral heavy-ion collisions the two nuclei geometrically \lq miss'
each other and no hadronic nucleon-nucleon collisions occur. At impact
parameters $b$ significantly larger than twice the nuclear radius
$R_\mathrm{A}$, the nuclei interact by photon exchange and photon-photon or
photon-Pomeron collisions \cite{upc_basics}.

We studied exclusive $\rho^0$ production at $\sqrt{s_\mathrm{NN}} =
130\,\mathrm{GeV}$ which, due to the coherent coupling of the exchanged photons
to the nuclei, have large cross sections.  Furthermore the final states are
restricted to low transverse momenta. The Au nuclei are not disrupted, and the
final state consists solely of the two nuclei and the vector meson decay
products:
$\mathrm{Au}+\mathrm{Au}\rightarrow\mathrm{Au}+\mathrm{Au}+\rho^0\rightarrow\mathrm{Au}+\mathrm{Au}+\pi^++\pi^-$.
In addition to coherent $\rho^0$ production, the exchange of virtual photons may
excite the nucei:
$\mathrm{Au}+\mathrm{Au}\rightarrow\mathrm{Au^\star}+\mathrm{Au^\star}+\rho^0\rightarrow\mathrm{Au^\star}+\mathrm{Au^\star}+\pi^++\pi^-$.

In general both types of reactions leave you with an \lq empty' detector, despite
of the decay products of the vector meson: $\rho^0\rightarrow\pi^++\pi^-$. These
two oppositely charged tracks are approximately back-to-back in the transverse
plane due to small $p_\mathrm{T}$ of the pair. The Au nuclei remain undetected
within the beam ($\mathrm{0n}, \mathrm{0n}$; no up- or downstream neutrons
emitted by the nuclei detected). Thus, for
$\mathrm{Au}\mathrm{Au}\rightarrow\mathrm{Au}\mathrm{Au}\rho^0$ we were
triggering on two single hits in opposite side quadrants of the CTB. Because
nearly all nuclear decays following photon absorption include neutron emission,
for $\mathrm{Au}\mathrm{Au}\rightarrow\mathrm{Au^\star}\mathrm{Au^\star}\rho^0$
the coincident detection of neutrons in the ZDCs was required $(\mathrm{xn},
\mathrm{xn})$.

\begin{figure}
\begin{center} 
\includegraphics[width=.36\textwidth]{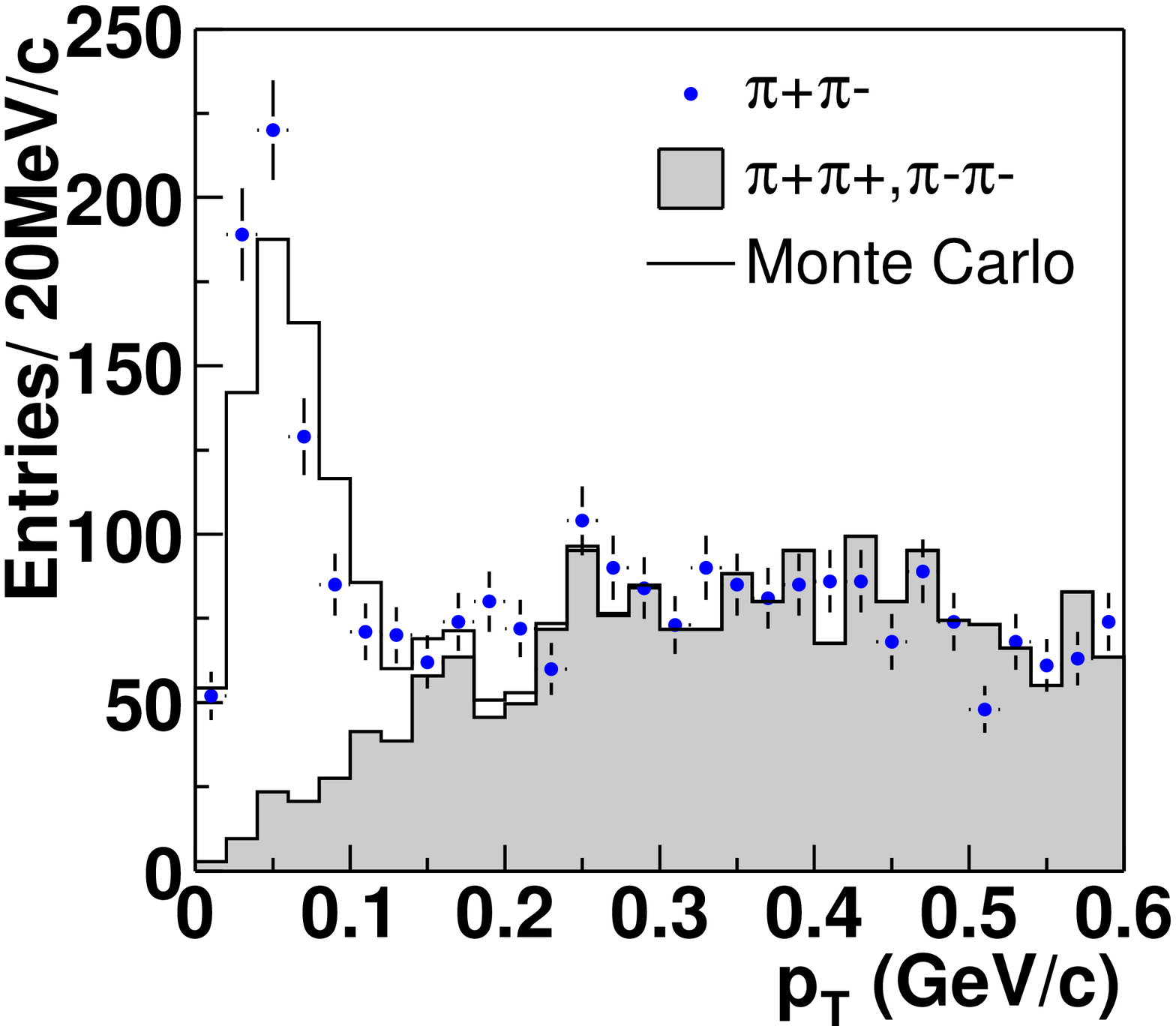}\hspace{.01\textwidth}
\includegraphics[width=.61\textwidth]{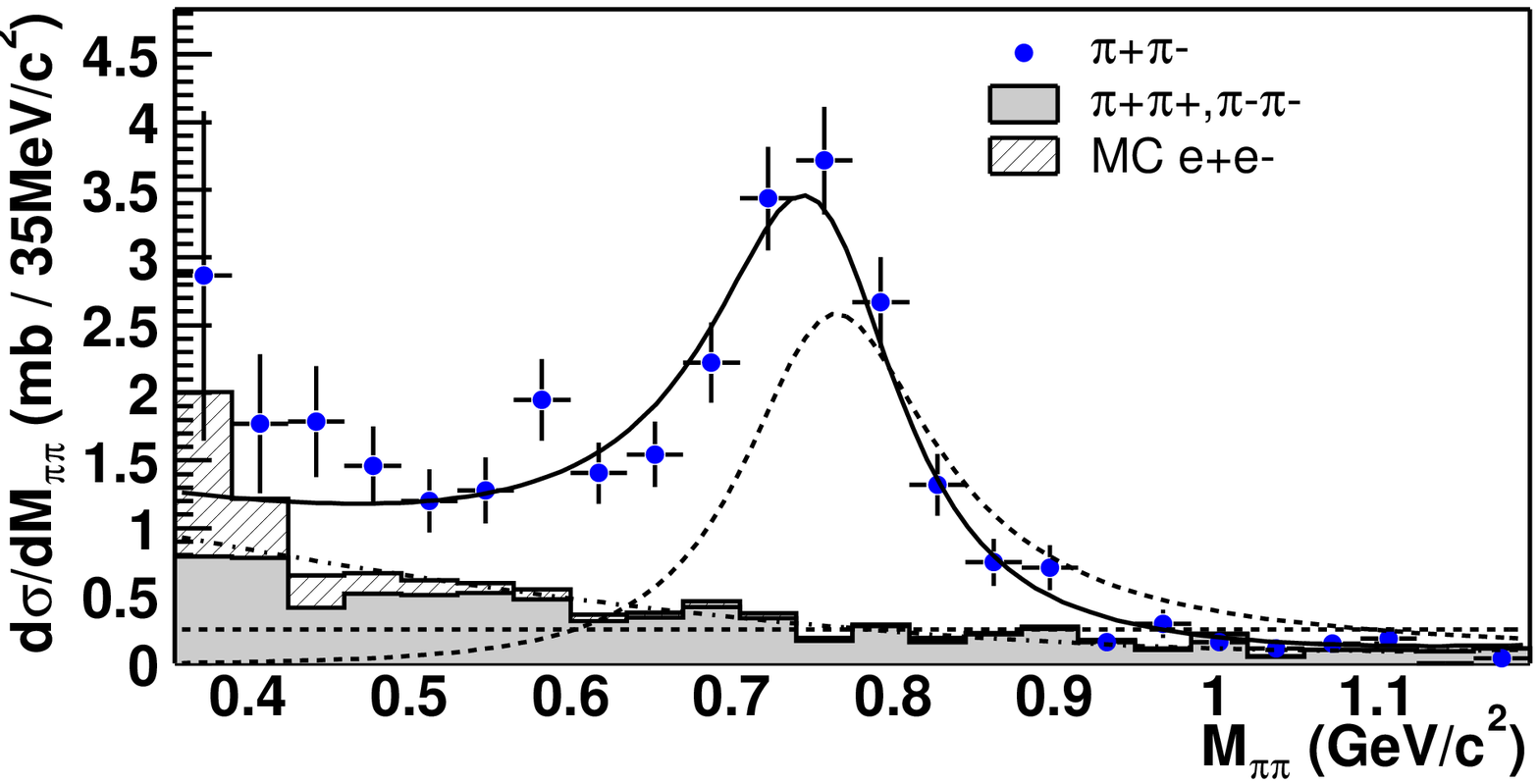}
\end{center}
\caption[Spectra of coherent $\rho^0$ production]{Spectra of coherent $\rho^0$
  production in 2-track events selected by the minimum bias trigger
  $(\mathrm{xn}, \mathrm{xn})$. In both figures points are oppositely charged
  pion pairs, and the shaded histograms are the normalized like-sign
  combinatorial background. {\bf (a)} $p_\mathrm{T}$ spectrum. {\bf (b)} $\D
  \sigma/\D M_{\pi\pi}$ for events with pair-$p_\mathrm{T} <
  150\,\mathrm{MeV/c}$. The hatched histogram contains an additional
  combinatorial backgroud contribution from coherent $\mathrm{e^+e^-}$ pairs}
\label{upc}
\end{figure}

The spectrum shown in Fig.~\ref{upc}~(a) shows the transverse momentum
distribution of pion pairs for the two-track event samples of the minimum bias
trigger $(\mathrm{xn}, \mathrm{xn})$. It is peaked at $p_\mathrm{T} \sim
50\,\mathrm{MeV/c}$ as is the $p_\mathrm{T}$ spectrum of the $(\mathrm{0n},
\mathrm{0n})$-triggered events (not shown here). This is the expected behavior
from coherent coupling.
 
The $\D \sigma/\D M_{\pi\pi}$ invariant mass spectrum for the $(\mathrm{xn},
\mathrm{xn})$ events with a pair-$p_\mathrm{T} < 150\,\mathrm{MeV/c}$ is shown
in Fig.~\ref{upc}~(b); the $(\mathrm{0n}, \mathrm{0n})$ events have a similar
$\D \sigma/\D M_{\pi\pi}$ spectrum. The spectrum is fitted by the sum of a
relativistic Breit--Wigner for $\rho^0$ production, the contribution of direct
$\pi^+\pi^-$ production and their interference. By extrapolating this fit to
full rapidity we measured the cross sections to
$\sigma(\mathrm{Au}\mathrm{Au}\rightarrow\mathrm{Au^\star_{\mathrm{xn},
    \mathrm{xn}}}\mathrm{Au^\star_{\mathrm{xn}, \mathrm{xn}}}\rho^0) =
28.3\pm2.0\pm6.3\,\mathrm{mb}$. This value is in agreement with theoretical
calculations \cite{WW,spencer}.  We estimate
$\sigma(\mathrm{Au}\mathrm{Au}\rightarrow\mathrm{Au}\mathrm{Au}\rho^0) =
370\pm170\pm80\,\mathrm{mb}$ $(0\mathrm{n}, 0\mathrm{n})$, see \cite{upc_paper}
for details.

\section{Conclusions and Outlook}
Nuclear matter created by collisions of heavy-ions is not a trivial convolution
of nuclear matter created by collisions of protons. Our analyses at 130 and
200\,GeV show that the fireball created at RHIC collisions is very dense and
possibly opaque. New energy loss mechanisms will be required to explain how
high-$p_\mathrm{T}$ particles interact with this hot and dense matter.

Our data show that nuclear matter produced at RHIC collisions is accurately
described by hydrodynamic models which assume local thermodynamic equilibrium at
very early times in the collision sequence. Away side jets appear to be missing
in central Au-Au collisions which suggests surface emission of jets or energy
loss in the partonic medium. The measurements of coherent $\rho^0$ production in
heavy-nuclei collisions confirm the existence of vector meson production in
those reactions and are in agreement with the theoretical expectations.

In summary the properties of nuclear matter at RHIC energies are not
inconsistent with local thermal equilibrium. First hints of the predicted jet
quenching effect are visible at RHIC. We are looking forward to extend our
studies during the upcoming run period which will serve us with up to 29~weeks
of d-Au (including cooldown) and 8~weeks of polarized proton collisions.

%


\begin{thebibliography}{15.}
\addcontentsline{toc}{section}{References}

\bibitem{rhic} RHIC Conceptual Design Report, BNL Report 52195 (1989)
\bibitem{drees} A.~Drees, private communication (2002)
\bibitem{tpc} H.~H.~Wieman \emph{et al.}, IEEE Trans.\ Nuc.\ Sci.~{\bf 44}, 671
  (1997); J.~H.~Thomas \emph{et al.}, Nucl.\ Inst.\ Meth.~{\bf A~478}, 166 (2002) 
\bibitem{ftpc} K.~H.~Ackermann, nucl-ex/0211014, to appear in a special
  volume of Nucl.\ Inst.\ Meth. dedicated to the accelerator and detectors at RHIC (2002)
\bibitem{flow} A.~M.~Poskanzer and S.~A.~Voloshin, Phys.\ Rev.~{\bf C~58},
  1671 (1998)
\bibitem{first_flow_paper} K.~H.~Ackermann \emph{et al.}, Phys.\ Rev.\ Lett.~{\bf 86}, 402 (2001)
\bibitem{mass_dependence_flow1}  C.~Adler \emph{et al.}, Phys.\ Rev.\ Lett.~{\bf
    87}, 182301 (2001)
\bibitem{mass_dependence_flow2}  C.~Adler \emph{et al.}, Phys.\ Rev.\ Lett.~{\bf
    89}, 132301 (2002)
\bibitem{hydro} P.~Huovinen, P.~F.~Kolb, U.~Heinz, P.~V.~Ruuskanen, S.~A.~Voloshin,
  Phys.\ Lett.~{\bf B~503}, 58 (2001)
\bibitem{jet_quenching_model} M.~Gyulassy, I.~Vitev, and X.-N.~Wang, Phys.\
  Rev.\ Lett.~{\bf 86}, 2537 (2001)
\bibitem{high_pt_flow} C.~Adler \emph{et al.}, nucl-ex/0206006, submitted to
  Phys.\ Rev.\ Lett.\ (2002)
\bibitem{high_pt} C.~Adler \emph{et al.}, nucl-ex/0210033, submitted to Phys.\ Rev.\ Lett.\ (2002)
\bibitem {upc_basics} G.~Baur, K.~Hencken and D.~Trautmann, J.\ Phys.~{\bf G~24},
  1657 (1998); C.~A.~Bertulani and G.~Baur, Phys.\ Rep.~{\bf 163}, 299 (1988)
\bibitem{WW} C.~F.~v.~Weizs\"acker, Z.\ Phys.~{\bf 88}, 612 (1934);
  E.~J.~Williams, Phys.\ Rev.~{\bf 45}, 729 (1934)
\bibitem{spencer} S.~Klein and J.~Nystrand, Phys.\ Rev.~{\bf C 60}, 014903
  (1999); A.~Baltz, S.~Klein and J.~Nystrand, Phys.\ Rev.\ Lett.~{\bf 89}, 012301 (2002)
\bibitem{upc_paper} C.~Adler \emph{et al.}, nucl-ex/0206004, accepted by Phys.\
  Rev.\ Lett.\ (2002)


\end{thebibliography}
\end{document}